\documentclass[unknownkeysallowed]{andp2012}
\usepackage[english]{babel}

\keywords{Many body localization; Long range interaction; Bethe lattice; Correlated energies}
\title{Localization and chaos in a quantum spin glass model in random longitudinal fields: Mapping to the localization problem in a Bethe lattice with a correlated disorder}
\author[F. Author]{Alexander Burin\inst{1,}\footnote{Corresponding author\quad E-mail:~\textsf{aburin@tulane.edu}}}
\address[1]{Tulane University, New Orleans, LA 70118, USA}
\shortauthors{A. Burin}
\begin{abstract}
The analytical solution of a many-body localization problem in a quantum   Sherrington-Kirkpatrick spin glass model in a random longitudinal field is proposed matching the problem with a model of Anderson localization in a Bethe lattice.
The localization transition is dramatically sensitive to the relationship between  interspin interaction and  random field revealing different regimes in which the interaction can either suppress or enhance the delocalization.  The localization is enhanced by  decreasing the temperature and the localization transition  shows a remarkable universality in a spin glass phase. The observed trends should be qualitatively relevant for other systems showing many-body localization. 
\end{abstract}
\shortabstract

\begin{document}
\maketitle

\section{Introduction}
\label{sec:intro}


Many-body localization has remained a primary focus of research for over a decade. Delocalized and localized regimes have qualitatively different thermodynamic properties. In the former case the whole system acts as a thermal bath for each small part of it\cite{Huse15Thermalization,Cohen13} while in the latter case different parts of the system are approximately independent and can be characterized by related integrals of motion \cite{Huse14IntMot}. The crossover between two regimes has been considered in different physical systems  \cite{Polkovnikov11RevModPhys} including quantum defects in a $^{4}$He crystal \cite{KaganMaksimov85}, anharmonic vibrations in polyatomic molecules \cite{Stewart83,LoganWolynes90,15LeitnerReview}, interacting electrons in Anderson insulators \cite{Gornyi05,Basko06} and quantum dots \cite{Silvan94Exp,MirlinQuantumDots16}, spin excitations in semiconductors and cold atomic systems  \cite{Lukin14MBLGen,Yao14MBLLongRange,Monro16,Bloch16MBLExp,LukinDiamond16} 
and in periodically driven systems \cite{Abanin15TimeDep,Lazarides15TDep,Yao16Floq}. Many body localization can be significant for quantum informatics \cite{Polkovnikov11RevModPhys,Yao16} because localization protects the quantum information while chaotic dynamics inevitably destroys it.

The many body localization (MBL) problem for interacting spins or particles can be formulated similarly to a single particle problem \cite{Anderson58}. The system Hamiltonian $\widehat{H}$ can be separated into an integrable static part $\widehat{H_{0}}$ and a dynamic perturbation $\lambda\widehat{V}$. The integrable Hamiltonian  $\widehat{H_{0}}$  has eigenstates represented by the products (or Slater determinants) of independent single particle or spin states of $N$ particles or spins with given population numbers $n_{i}$ or spin projections $S_{i}^{z}$ to the quantization axis $z$. A perturbation $\lambda\widehat{V}$ causes the transitions between those states. The quasistatic Hamiltonian can be further separated into a pure random field and an interparticle interaction (e. g. binary spin-spin interaction) which is responsible for the ``many-body" nature of the problem.

The integrability of a zeroth order problem ($\widehat{H_{0}}$) can be expressed in terms of integrals of motion \cite{Huse14IntMot} (population numbers $n_{i}$ or spin projections $\sigma_{i}^{z}$ for a zeroth order problem). These integrals of motion should be weakly disturbed  when $\lambda \ll 1$. 
The delocalization takes place at sufficiently large perturbation $\lambda>\lambda_{c}$. Delocalized eigenstates are represented by combinations of a macroscopic number of zeroth order product states which is comparable to the total number of states uniformly spread over the phase (Fock) space. The phase space intersections of  delocalized states should lead to their level repulsion and consequently a Wigner Dyson statistics of their energy levels. This statistics is considered as a signature of chaotic and ergodic behaviors in a quantum many-body problem \cite{Cohen13} similarly to a single particle problem \cite{ShklovskiiShapiro93}. 

\begin{figure}[h!]
\centering
\includegraphics[width=\columnwidth]{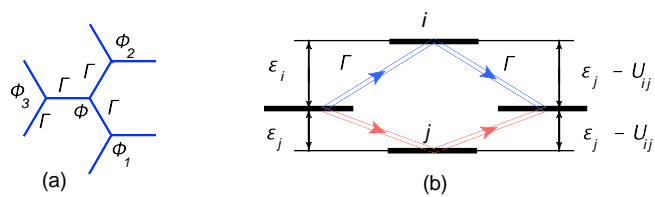}
\caption{\small (a) A Bethe lattice model with the coordination number $N=3$, site energies are represented by random energies $\Phi$, $\Phi_{1}$,.. and the coupling between sites is determined by the transverse field  $\Gamma$. (b) Interference of two subsequent spin flip transitions $(i, j)$ and $(j, i)$.}
\label{fig:Interf}
\end{figure}

An MBL problem is yet more complicated compared to a single particle problem. It is very difficult for numerical studies because a number of many-body states   grows exponentially with the system size \cite{FleishmanAnderson80,AltshullerGefen97}.  Therefore analytical solutions are very significant for understanding MBL.  The exact analytical solution exists for the localization problem on a Bethe lattice  (see Refs. \cite{AbouChacra73,Shapiro83,Efetov83Rev,Mirlin91} and Fig. \ref{fig:Interf}.a representing a Bethe lattice) and this solution can possibly describe many-body localization because of similarities in the phase spaces  \cite{AltshullerGefen97,Gornyi05,Basko06}. 



It turns out that a reasonably accurate matching to the Bethe lattice problem can be attained in the presence of a strong diagonal interaction between   particles or spins as it was shown in a quantum random energy spin glass model \cite{Laumann14,Laumann16}. Many-body localization in this  model  can be almost precisely described by the solution of the matching Bethe lattice localization problem.

However, a random energy model has certain features of a single particle problem which are not typical for a many-body problem. In that model a single spin flip leads to a change in total energy comparable to that energy itself. This is the consequence of the lack of correlations between diagonal energies for all states which significantly simplifies the analytical solution of the problem. Such behaviour is typical for a single particle problem while in a large many-body  system the energy change associated with a single particle excitation is  much less than the energy of the system itself. Because of this difference the localization transition  in a random energy model can be dramatically sensitive to a system energy and temperature which might not take place in more realistic settings.  A quantum random energy model contains only two parameters including interspin interaction and transverse field, responsible for spin flips, while the typical MBL problem contains static disorder, many-body and dynamic interactions. These shortcomings limit the applicability of analytical results for random energy model  to the MBL problem. In this work one more step towards analytical solution of the realistic MBL problem  is suggested considering this  problem in a quantum Sherrigton Kirkpatrick (SK) spin glass model \cite{Sherrington75} in a random longitudinal field. 

The advantages of this model include a finite energy change associated with a single spin flip which makes it closer to realistic systems. The problem can still be resolved analytically at finite and infinite temperatures in spite of strong correlations between energies of coupled states. Moreover the obtained solutions identify and separate effects of disordering and interaction on the MBL transition and the obtained parametric dependencies for that transition should be transferable to a large extent to more realistic problems with a short range interaction.  The  SK model still has shortcomings including an infinite-range interaction and consequently vanishing of localization threshold in a thermodynamic limit.  However, the transition is expected to get sharper with increasing the system size and therefore the theory is still potentially applicable to other MBL transitions. 

Although in the case of an infinite range of interaction the concepts of space or location are not applicable to a system dynamics, it is still relevant for a variety of physical systems including for instance delocalized electrons in quantum dots \cite{Silvan94Exp,MirlinQuantumDots16}. The localization-delocalization transition is considered in the Fock space of independent particle product states and it describes the transition between non-ergodic and chaotic  (ergodic) behaviors. 
While the infinite range interaction is hard to realize in real spin systems it is possible to use trapped cold atoms to create a similar setup \cite{Lukin14MBLGen,Yao14MBLLongRange,Monro16} and the theoretical predictions of the present work can be hopefully tested in those systems. The theory is  extendable  to short-range interactions as demonstrated in Sec. \ref{sec:PowLaw}.

The paper is organized as follows. The model is introduced in Sec. \ref{sec:model}.  The matching Bethe problem and its solution are described in Sec. \ref{sec:Bethe}. In Sec. \ref{sec:SGinfT} the main results of the work related to the MBL transition in a spin glass model at infinite temperature are described. In Sec.  \ref{Sec:T}  these results are extended to finite temperatures.  Another extension of the theory to the spin glass with a  power law interaction $1/r^{\alpha}$ is discussed in Sec. \ref{sec:PowLaw}  showing its consistency with previous work \cite{ab06preprint,ab15MBL,Yao14MBLLongRange}. The conclusions are given at the end of the work in Sec. \ref{sec:Concl}.
Derivations can be found in the Supporting Information according to the references in the main text.

\section{Model}
\label{sec:model}


The Sherrington-Kirkpatrick spin glass model \cite{Sherrington75} with a random binary interaction of infinite radius in a small transverse field and random longitudinal field is considered. The Hamiltonian of $N$ interacting  spins $1/2$ represented by Pauli matrices $\sigma_{i}^{\mu}$ ($i=1,..N$, $\mu=x, y, z$) in a transverse field $\Gamma$ reads 
\begin{eqnarray}
\widehat{H}=\widehat{H}_{0}+\widehat{V},~
\widehat{H}_{0}=\sum_{i=1}^{N}\phi_{i}\sigma_{i}^{z} + \sum_{i=1}^{N}\sum_{j=1}^{i-1}J_{ij}\sigma_{i}^{z}\sigma_{j}^{z}, 
\nonumber\\
 \widehat{V}=\Gamma\sum_{i}\sigma_{i}^{x}. 
\label{eq:H}
\end{eqnarray}

The problem has three different energy scales associated with random fields $\phi_{i}$, interactions $J_{ij}$ and a transverse field $\Gamma$. Random fields are  assumed to be independent, having zero average and possessing a Gaussian distribution with the width $W$. Interactions $J_{ij}$ are assumed to be independent and  distributed according to the Gaussian law with a zero average and the root mean square $J/\sqrt{N}$ \cite{Sherrington75} (except for Sec. \ref{sec:PowLaw}). The typical binary interaction will be also denoted as $U = 4J/\sqrt{N}$ for the convenience. The model is integrable  at  a zero transverse field  and/or in the absence of interactions.  

The work main target is to determine the localization threshold which can be expressed as the maximum transverse field $\Gamma_{c}$ where the system remains quasi-integrable or localized. Below, the results for this localization threshold are derived and compared to each other for the matching Bethe lattice problem and the model of interest, Eq. (\ref{eq:H}).

\section{Matching Bethe lattice problem}
\label{sec:Bethe}

The matching Bethe lattice problem for the spin glass model at an infinite temperature can be formulated as following. Each eigenstate $|a>$ of the Ising model (Eq. (\ref{eq:H}) at $\Gamma=0$) can be characterized by the sequence of $N$ spin projections to the $z-$axis $a=\{\sigma_{ia}=\pm 1\}$ ($i=1,...N$). It corresponds to a single site in the Bethe lattice graph (see Fig. \ref{fig:Interf} (a)). Each site is connected by the transverse field $\Gamma$ to $N$ other sites $a_{k}$ ($k=1,...N$) different from the given state by a one spin flip ($\sigma_{k}^{z}=-\sigma_{ka}$), each site can be characterized by the diagonal energy $\Phi_{a}=\sum_{i}\phi_{ia}\sigma_{ia} + \frac{1}{2}\sum_{i,j}J_{ij}\sigma_{ia}\sigma_{ja}$.

It turns out that this problem can be resolved in the limit of a large coordination number $N \gg 1$ where spin-flip transition energies 
\begin{eqnarray}
\epsilon_{i}=2\sigma_{ia}(\phi_{i}+\sum_{k}J_{ik}\sigma_{ka})
\label{eq:SpFlipE}
\end{eqnarray}
can be considered as approximately independent (see Supporting Information, Sec. \ref{sec:SI1}), averaging to zero and distributed according to the Gaussian law 
\begin{eqnarray}
p(\epsilon)=\frac{1}{\sqrt{2\pi}W_{1}}e^{-\frac{\epsilon^2}{2W_{1}^2}}, ~ W_{1}=2\sqrt{W^2+J^2}. 
\label{eq:SpFlipEDistr}
\end{eqnarray}

Delocalization is determined by sequences of spin flip transitions which can be described within the forward approximation \cite{AltshullerGefen97,Laumann14,Laumann16} equivalent to the self-consistent theory of localization \cite{Anderson58,AbouChacra73}. The $n^{th}$ order forward approximation is determined by the sequence of $n$ spin flips characterized by $n$ transition energies for one, two and more spin flips  ($\epsilon_{1}$, $\epsilon_{12}$,... $\epsilon_{12..n}$). The contribution of this process to the forward approximation can be estimated with the logarithmic accuracy as 
\begin{eqnarray}
X_{n} \approx P_{n}(0, 0...)\left[N\Gamma\ln\left(\frac{W_{1}}{\Gamma}\right)\right]^{n}
\label{eq:BethForw}
\end{eqnarray}
where $P_{n}(0, 0,...)$ expresses a joint probability that all   energy changes ($\epsilon_{1}$, $\epsilon_{12}$,... $\epsilon_{12..n}$) are equal to zero (cf. \cite{Laumann14}). In  the case of a relatively small number of spin flips, $n \ll N$,  these energies can be approximated by $\epsilon_{1,..k}=\sum_{i=1}^{k}\epsilon_{i}$ where $\epsilon_{i}$ is the energy of $i^{th}$ spin flip. This contrasts to  a random energy model where all energies for $k$ spin flips are determined by energy differences $\Phi_{k}-\Phi_{0}$ representing final and initial states respectively and all energies $\Phi_{k}$ are independent of each other. 

 Assuming that energies of different spin flips, $\epsilon_{k}$ are approximately independent of each other and characterized by distributions Eq. (\ref{eq:SpFlipEDistr}) one can evaluate the probability $P_{n}(0, 0...)$ as 
\begin{eqnarray}
P_{n}(0, 0,...)\approx \prod_{i=1}^{n}\int_{-\infty}^{\infty}d\epsilon_{k} p(\epsilon_{k})
\delta\left( \sum_{i=1}^{k} \epsilon_{k}\right) =p(0)^{n}, 
\label{eq:SpFlipEDistr1}
\end{eqnarray}
so it can be replaced with the product of $n$ independent probabilities as in the problem with uncorrelated energies \cite{Anderson58,AbouChacra73,Laumann14}. 
The similar result has been earlier derived for electrons in quantum dots in Ref. \cite{Ros15}. The more accurate analysis given in the Supporting Information, Sec. \ref{sec:SI1},  confirms the validity of decoupling of the probability $P_{n}$ into the product of $n$ independent factors $p(0)$ in the Sherrington-Kirkpatrick spin glass model at an infinite temperature, while at a finite temperature exceeding the glass transition temperature these correlations can slightly suppress the localization (see Sec. \ref{Sec:T}). 

The localization transition can be determined using the divergence of a forward approximation, Eq (\ref{eq:BethForw}), at large $n$. Using Eq. (\ref{eq:SpFlipEDistr1}) one can evaluate the localization threshold in terms of the critical transverse field $\Gamma_{c}$ as
\begin{eqnarray}
\Gamma_{c,B}=\frac{1}{4p(0)N\ln(1/(p(0)\Gamma_{c}))}\approx \frac{1}{4p(0)N\ln(N)}.  
\label{eq:abThr}
\end{eqnarray}
The more accurate derivation of the localization threshold for a Bethe lattice problem is given in the Supporting Information, Sec. \ref{sec:BLSol}. 

The critical transverse field $\Gamma_{c}$ given by Eq. (\ref{eq:abThr}) vanishes in the thermodynamic limit of an infinite system ($N \rightarrow \infty$) as $(N\ln(N))^{-1}$ in contrast with the localization threshold in a random energy model at a finite temperature that remains finite \cite{Laumann14}. I believe that this regime is still important since the same trend takes place in many systems of interest including interacting electrons in quantum dots where the localization threshold is determined by interlevel splitting vanishing in the thermodynamic limit \cite{AltshullerGefen97,Gornyi05,abGorniyMirlinDot}.   Moreover one can expect the narrowing of the relative width of the localization transition $\Delta\Gamma_{c}/\Gamma_{c}$ to zero with increasing $N$. Indeed, the delocalization in the Bethe lattice problem can be associated with the divergence of the forward approximation which can be expressed by the infinite power series with $n^{th}$ order term behaving as $(\Gamma/\Gamma_{c})^{n}$ \cite{Anderson58,Mirlin91}. In the finite system the maximum contribution is given by $n \sim N$ and it scales as $(\Gamma/\Gamma_{c})^{N}$. The exponential growth in the number of resonances indicates that delocalization begins at $\Gamma = \Gamma_{c}$ and the number of resonances substantially exceeds unity at $\Delta\Gamma = \Gamma-\Gamma_{c} \sim \Gamma_{c}/N$. Thus the finite size width of the transition  is expected to scale as $\Gamma_{c}/N$ and a relative transition width $\Delta\Gamma/\Gamma_{c}$ approaches zero within the thermodynamic limit. Therefore one can still talk about the localization transition in spite of a vanishing localization threshold in a thermodynamic limit. 

\section{Spin glass model}
\label{sec:SGinfT}

The consideration begins with the infinite temperature case where the transition energy distribution has a Gaussian form Eq. (\ref{eq:SpFlipEDistr}) and any correlations between different transition energies can be neglected. 
In contrast to a random energy model \cite{Laumann14} the localization criterion Eq. (\ref{eq:abThr}) obtained for the Bethe lattice problem cannot be transferred directly to the model Eq. (\ref{eq:H}) because it does not distinguish between spin-spin interactions and random fields. This is the consequence of the lack of interference between different paths in the Bethe lattice problem relative to that which takes place in the spin glass model (see Fig. \ref{fig:Interf}. b). 

The interference takes place because the spin flips occurring in different orders lead to the same final state (see Fig. \ref{fig:Interf} (b) for two spins) while different sequences of transitions in the Bethe lattice lead to different states. $n$ transitions from the given product state in the spin glass model corresponds to $n$ spin flips leading to ${N \choose n} = N!/(n!(N-n)!)$ different states in contrast to $N^{n}$ states for the Bethe lattice. There are $N!/(N-n)! \approx N^{n}$ different sequences corresponding to $n$ spin flips and for each sequence there are $n!$ interfering paths corresponding to $n$ spin permutations all leading to the same state.

The significance of such interference can be illustrated considering flips of two spins $a$ and $b$ characterized by spin flip energies $\epsilon_{i}$ and $\epsilon_{j}$ (see Fig. \ref{fig:Interf} (b)) occurring in a different order. The matrix element of the transition from the initial state to the state with both flipped spins $i$ and $j$ can be evaluated using second order perturbation theory as the sum of two path contributions
\begin{eqnarray}
V_{ij}=\frac{\Gamma^2}{\epsilon_{i}}+\frac{\Gamma^2}{\epsilon_{j}}=\frac{\Gamma^2(\epsilon_{i}+\epsilon_{j})}{\epsilon_{i}\epsilon_{j}}. 
\label{eq:Vab}
\end{eqnarray}
The perturbation is most efficient under resonant conditions where the total energy change $\epsilon_{ij}$ after two flips approaches zero. This energy can be expressed as $\epsilon_{ij}=\epsilon_{i}+\epsilon_{j}-U_{ij}$ where the interaction term $U_{ij}$ is defined as $U_{ij} =4J_{ij}\sigma_{i}^{z}\sigma_{j}^{z}$ and $\sigma_{i}^{z}, \sigma_{j}^{z}$ are spin projections in the initial state \cite{Efros75}. In the case of a weak interaction $U_{ij} < \epsilon_{i}, \epsilon_{j}$ a destructive interference suppresses second order transitions. 

A resonant interaction takes place for spin flip energies $\epsilon_{i}$ or $\epsilon_{j} \sim \Gamma$. Consequently interference does not affect resonances for two spin transitions if the spin-spin interaction $U=4J/\sqrt{N}$ (here and below the notation $U$ stands for the typical binary spin-spin interaction) exceeds a resonant energy given by the transverse field $\Gamma$. In the case of interest   $\Gamma \sim \Gamma_{c}$ this field is given by the characteristic minimum spin flip energy per the state, $\Gamma_{c} \sim 1/(p(0)N)\sim W_{1}/N$ (see Eq. (\ref{eq:abThr})). The destructive interference can be ignored in the case of the {\it strong interaction} defined as 
\begin{eqnarray}
1 \ll 4\sqrt{N}p(0)J=Np(0)U \sim \frac{U}{\Gamma_{c}}.   
\label{eq:n2StrInt0}
\end{eqnarray}
The {\it weak interaction} case takes place in the opposite regime of $U \ll 1/(Np(0))$. Below these  two regimes are treated separately. 

\subsection{Strong Interaction}

Here the localization in the strong interaction regime at an infinite temperature is considered at a semi-qualitative level, while more accurate derivation using a self-consistent forward approximation can be found in the Supporting Information, Sec. \ref{sec:SIStrong}.  Qualitatively the strong interaction regime can be defined as following. Assume that the characteristic random field $W$ exceeds the interaction strength $J\sim U\sqrt{N}$. Then one can remove spins with large spin flip energies $\epsilon > \xi W$ from the consideration assuming them to be slow compared to other spins. The number of remaining spins decreases with the cutoff parameter $\xi$ as $N_{\xi}=\xi N$ while their binary interaction $U$ remains unchangeable.  Choosing $\xi \approx (J/W)^2$ one ends up with $N_{min}=NJ^2/W^2=(NU/W)^2$ remaining spins characterized by the random field strength $W_{min} \sim J^2/W=NU^2/W$ equal to the interaction strength $J_{min}=U\sqrt{N_{min}}$. One should notice that the number of remaining spins, 
\begin{eqnarray}
N_{min}=N\frac{J^2}{W^2},
\label{eq:ScalMinN}
\end{eqnarray}
is much greater than unity in the strong interaction regime and this regime can be thus reduced to the spin glass problem with a small random field. Particularly in the strong interaction regime the spin glass transition temperature is given by $k_{B}T_{f} \sim J^2/W$ \cite{Hadjiagapiou14}  while there is no spin glass transition in the weak interaction limit where $N_{min} \ll 1$. 

Below the results of the $n^{th}$ order forward approximation are introduced for the strong interaction regime, then the applicability limits for the forward approximation are established and finally the expression for localization threshold is derived. 

\subsubsection{An $n^{th}$ order forward approximation.} 
\label{subsec:forwn}

For $n$ spin flips determining the $n^{th}$ order forward approximation the scale of $n-$spin interaction can be estimated as $U_{n} \approx U\sqrt{n}$. This energy gives the upper limit for the maximum spin energy where destructive interference is still avoided (cf. Eq. (\ref{eq:Vab})) suggesting that all $n!$ spin flip permutations for the given spin sequence contribute to independent resonant interactions. Consequently, the maximum energy in the logarithm in the definition of the localization threshold for the Bethe lattice in Eq. (\ref{eq:abThr}) should be replaced with $U_{n}$ leading to the estimate of the $n^{th}$ order contribution to the forward approximation in the form
\begin{eqnarray}
X_{n}\sim \frac{N!}{(N-n)!N^{n}}\left(4N\Gamma p(0)\ln\left[\frac{U\sqrt{n}}{\Gamma}\right]\right)^{n}. 
\label{eq:nth_order}
\end{eqnarray}

The delocalization threshold in the $n^{th}$ order forward approximation can be found setting $X_{n} \approx 1$. This yields 
\begin{eqnarray}
\Gamma_{cn} \approx \frac{\eta_{n}}{4Np(0)\ln\left[p(0)NU\sqrt{n}\right]}, \eta_{n}=\left(\frac{(N-n)!N^{n}}{N!}\right)^{\frac{1}{n}}
\label{eq:Gamnth}
\end{eqnarray}
The prefactor $\eta_{n}$ changes between $1$ for $n \ll N$ and $e$ for $n \rightarrow N$.  These results are applicable until the forward approximation is valid. It fails for very large $n$ as discussed below. 

\subsubsection{Validity of a forward approximation}

The forward approximation skips spin flip sequences involving several flips of the same spin. This can be formally justified by the large coordination number $N$ which is true for a small number $n$ of spin flips. The involvement of a new spin in the $n+1^{st}$ step occurs with the probability proportional to the number of available spins ($N-n$) while there are around $n$ choices for the second flip of one of already used spins. However the energies of remaining $N-n$ spins are distributed within  the energy domain $(-W, W)$  while the flipped spin energies do not exceed the $n-$spin interaction $U_{n} =U\sqrt{n} \ll W$ which makes the resonance probability greater by the factor $W_{1}/U_{n}$ for the ``backwards" process. Consequently, multiple flips of the same spin are significant  in the case 
\begin{eqnarray}
 n >  U^2 N/W_{1}^2 \sim  N_{min}. 
\label{eq:ForwLimit}
\end{eqnarray}
Thus the forward approximation is valid only for $n<N_{min}$ (cf. the analysis of interacting resonances in Ref. \cite{abGorniyMirlinDot}). This is also demonstrated in a more quantitative manner using the Green function method in the Supporting Information, Sec. \ref{subsec:ForwSI}. 

According to Eq. (\ref{eq:ForwLimit}) in the case of a small random longitudinal field, $W \leq J=U\sqrt{N}$, the forward approximation is valid until the number of spin flips approaches the total number of spins ($N_{min} \sim N$).  In the opposite case the maximum number of spin flips is limited to the number of spins $N_{min}$ obtained for the rescaled problem with the interaction comparable to a random field, Eq. (\ref{eq:ScalMinN}). Below the localization in the model with weak random field is considered first and then the results are extended  to stronger randomness. 

\subsubsection{Localization threshold for small random field} 
\label{subsec:strsm}

In the case of a small random field, $W \leq J$, the forward approximation is valid until the number of spin flips is less than the total number of spins, $n < N$. The localization threshold can be estimated using Eq. (\ref{eq:Gamnth}) either considering its minimum for $\Gamma_{cn}$ with respect to $n$ realized at  $n\sim N/\ln(N)$  or exploiting the limit $n \sim N$ where the forward approximation is still approximately valid. The first estimate coincides with the logarithmic accuracy with the estimate for the equivalent Bethe lattice problem, Eq. (\ref{eq:abThr}), while the second estimate can exceed it by a factor  $\eta_{n}$  ranging between $1$ and $e$ (see Eq. (\ref{eq:Gamnth})). It is originated from the reduction of the number of system pathways expressed by the factorial term for large $n\sim N$. 

The present method cannot distinguish between these two estimates since the theory does not go beyond the forward approximation. Although a substantial delocalization in the Fock space involving practically all spin flips can already be expected  for $\Gamma > \Gamma_{c,B}$ (Eq. (\ref{eq:abThr})), it can be insufficient to ensure a truly chaotic dynamics characterized by the Wigner Dyson energy level statistics \cite{Cohen13} which does not necesserily takes place in the Bethe lattice problem \cite{DeLuca14}. A true ergodicity might need the interference of many resonant paths which requires $n \sim N$, where some increase of the localization threshold compared to the Bethe lattice estimate in Eq. (\ref{eq:abThr}) can be expected according to Eq. (\ref{eq:Gamnth}). 

The latter expectation is consistent with numerical studies of localization transition in high dimensions \cite{GarciaCuevas07} using the Wigner Dyson statistics as the delocalization criterion. These studies show that the localization threshold behaves in a qualitatively similar manner to the predictions for the Bethe lattice but exceeds the related theory predictions by the factor of $2$ possibly  due to the parameter $\eta_{n}$ in Eq. (\ref{eq:Gamnth}). 

Consequently, the present theory can estimate the localization threshold within  accuracy of the unknown factor $\eta \sim 1$ ($1 < \eta \leq e$)  as 
\begin{eqnarray}
\Gamma_{c} \approx \frac{\eta}{4Np(0)\ln(N)}.  
\label{eq:GamSK}
\end{eqnarray} 
The factor $\eta$ can be determined in future numerical and/or experimental studies.  I will use $\eta=2$ in the future discussions in accord with the numerical studies \cite{GarciaCuevas07}.

\subsubsection{Large random field} 
\label{subsec:strl}

In the case of a large random field compared to the spin-spin interaction, $W > J$, the $n^{th}$ order forward approximation is valid in a sufficiently small order $n$, such that $n<N_{min}\approx N(J/W)^2$, Eq. (\ref{eq:ForwLimit}). Consequently the localization threshold expression given by Eq. (\ref{eq:Gamnth}) at $n=N_{min}$ given by $\Gamma_{c}=1/(4Np(0)\ln(N_{min}))$ can serve as a lower estimate for the threshold. 

An upper estimate can be obtained considering the localization in the rescaled problem where only spins with spin flip energies comparable or less than $J^2/W$ are left as described in the beginning of this section. This case can be described by Eq. (\ref{eq:GamSK}) with the reduced number of spins $N_{min}$ given by Eq. (\ref{eq:ScalMinN}).   The delocalization in the resonant subsystem of $N_{min}\gg 1$ spins should be sufficient to stimulate the irreversible dynamics of all remaining spins which interact with this resonant subsystem as with an ergodic bath. Considerations of the similar problems in Refs. \cite{Huse15Bath,abGorniyMirlinDot} for particle transitions induced by the energy exchange with the bath are completely applicable here taking the advantage of the fact that the bath level splitting decreases exponentially with the number of spins. Consequently one can describe the localization using  Eq. (\ref{eq:GamSK}) with the modified logarithmic factor as 
\begin{eqnarray}
\Gamma_{c,str} \approx \frac{\eta}{8Np(0)\ln\left(\frac{U}{\Gamma_{c,str}}\right)}\approx \frac{\eta}{4Np(0)\ln(N_{min})}.  
\label{eq:StrIntAns}
\end{eqnarray}
leaving a small uncertainty in the numerical constant factor $\eta$ that should not exceed  $2$ (cf. \cite{GarciaCuevas07}).  

The case of a weak interaction cannot be reduced to the model with a small random field. The delocalization in this case takes place at substantially larger transverse field compared to  $W/N$ due to the pairwise interactions of spins as described below. 

\subsection{Weak interaction}
\label{subsec:weak}

In the weak interaction regime spin-spin interactions $U$ are smaller then a transverse field $\Gamma$. One can then restrict the consideration to spins with flip energies smaller than $\xi \Gamma$ ($\xi$ is a scaling parameter, $\xi >1$) assuming  other spins to be slow. Then the number of remaining spins scales as $N_{\xi}=N\xi\Gamma/W$ and their interaction scales as $J\sqrt{\xi\Gamma/W}= \Gamma \sqrt{\xi \frac{W^2}{N^2 U \Gamma} \left(\frac{NU}{W}\right)^3}$ which is smaller then the transverse coupling $\Gamma$. Indeed the scaling factor of $\xi$ is of order of unity, the second factor under the square root  is of order of unity near the localization threshold given by Eq. (\ref{eq:WeakIntAns}) and the third factor is small in the weak interaction limit opposite to  Eq. (\ref{eq:n2StrInt0}). Therefore the spin-spin interaction is less than the transverse field and one can approximately diagonalize each individual spin Hamiltonian in its transverse and longitudinal fields. The equivalent Hamiltonian can be set in the form  \cite{ab06preprint}
\begin{eqnarray}
\widehat{H}_{weak}= \sum_{i=1}^{N_{\xi}}E_{i}\sigma_{i}^{z}
\nonumber\\
+\frac{1}{2}\sum_{ij}^{N_{\xi}}J_{ij}\left(\frac{\epsilon_{i}}{E_{i}}\sigma_{i}^{z} - \frac{\Gamma}{E_{i}}\sigma_{i}^{x}\right)\left(\frac{\epsilon_{j}}{E_{j}}\sigma_{j}^{z} - \frac{\Gamma}{E_{j}}\sigma_{j}^{x}\right), 
\nonumber\\
E_{i}=\sqrt{\epsilon_{i}^2+\Gamma^2}. 
\label{eq:Hweak}
\end{eqnarray}
Here energies $\epsilon_{i}$ represent independent spin flip energies distributed uniformly in the domain $(-\xi\Gamma, \xi\Gamma)$.

The delocalization is associated with the flip-flop terms $\sigma_{i}^{x}\sigma_{j}^x$ while the terms $\sigma_{i}^{x}\sigma_{j}^z$ are not significant because  single spin level splittings exceed  $\Gamma$. In the case of $\sigma_{i}^{x}\sigma_{j}^x$ interaction there are flip-flop transitions characterized by an energy change equal to a two spin flip energy difference, which can be made arbitrarily small \cite{ab98book,ab98prl}.

Each product state with fixed spin projections  is coupled to $N_{\xi}^2/2$ other states with the coupling strength $U$ ($\xi \sim 1$). Consequently delocalization takes place at around one resonant interaction per state, i. e. $N_{\xi}^2 U/\Gamma  \sim 1$. The conservative estimate using the matching Bethe lattice problem with logarithmic factor determined similarly to the case of a strong interaction  results in the expression (see Supporting Information, Sec. \ref{sec:SIWeak})
\begin{eqnarray}
\Gamma_{c}=\frac{\eta_{1}}{7 N^{2} p(0)^2 U}, ~ \eta_{1} \approx 1. 
\label{eq:WeakIntAns}
\end{eqnarray}
There is no logarithmic factor in the localization threshold expression  in contrast to Eq. (\ref{eq:StrIntAns}) because the interaction is comparable to the transition amplitudes in Eq. (\ref{eq:Hweak}).

\begin{figure}[h!]
\centering
\includegraphics[width=\columnwidth]{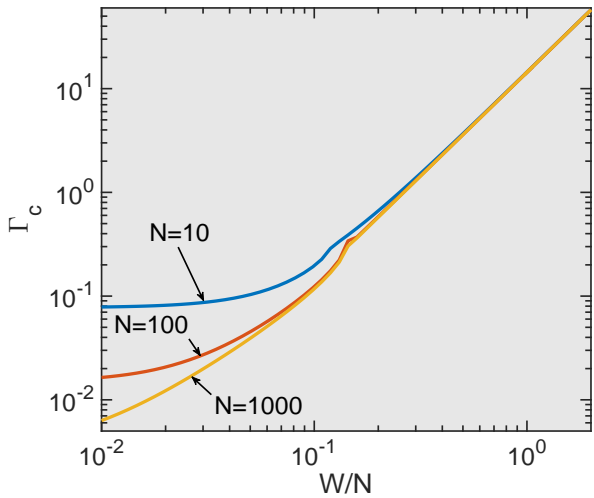}
\caption{\small Localization threshold dependence on a relative random field strength (Eq. (\ref{eq:FullAns}).}
\label{fig:ThrInfW}
\end{figure}

The results are valid until $\Gamma_{c} \ll W$. In the opposite case $\Gamma_{c} \gg W$ corresponding to a very strong random field, $W \gg JN^{3/2}$, off-diagonal interaction dominates making the problem similar to the $X-Y$ model. In this regime I would expect a stronger increase of the threshold field $\Gamma_{c}$ with disordering strength $W$ then in Eq. (\ref{eq:WeakIntAns}) because of the specific of delocalization in the XY model \cite{ab15MBLXY}. The analysis of this regime in detail is beyond the scope of the present work.

\subsection{Discussion}

\begin{figure}[h!]
\centering
\includegraphics[width=\columnwidth]{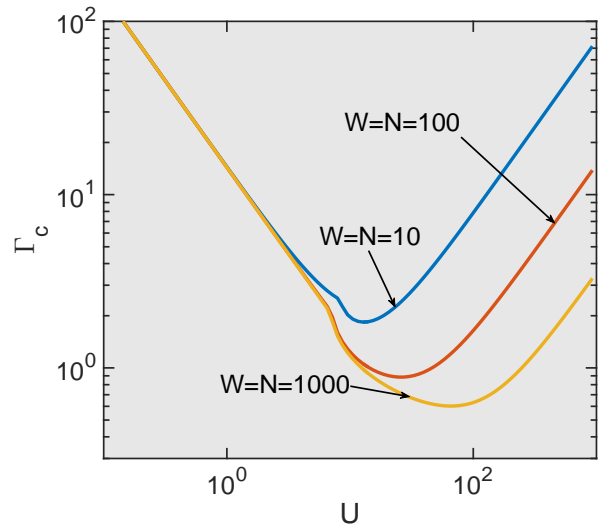}
\caption{\small Localization threshold dependence on an interspin interaction (Eq. (\ref{eq:FullAns})).}
\label{fig:ThrInf}
\end{figure}

Thus at the infinite temperature the localization threshold behavior can be described by Eqs. (\ref{eq:StrIntAns}), (\ref{eq:WeakIntAns}) in the regimes of weak and strong interactions, respectively. It is convenient to express dependencies of $\Gamma_{c}$ on the random field strength $W$ and interaction $U$ separately to examine the localization threshold dependence on these parameters. Then the results in Eqs.  (\ref{eq:StrIntAns}), (\ref{eq:WeakIntAns})  can be presented together  as
\begin{eqnarray}
\Gamma_{c}
\approx\begin{cases}
\frac{W}{N}\frac{\sqrt{\pi}}{\sqrt{2}\ln\left(\frac{U}{\Gamma_{c}}\right)}\sqrt{1+\frac{J^2}{W^2}}, ~ U \gg \frac{W}{N},\\
\frac{1}{U}\left[\frac{W}{N}\right]^{2}\frac{32\pi}{7},
   ~ U \ll \frac{W}{N}. 
  \end{cases}  
\label{eq:FullAns}
\end{eqnarray} 
where a random field parameter $W$ enters together with the number of spins $N$ in the combination $W/N$ estimating a typical minimum field for one out of $N$ spins. This minimum field  is most suitable for the resonant interaction.  The unknown numerical factors are chosen as $\eta=2$ (cf. Ref. \cite{GarciaCuevas07}) and $\eta_{1}=1$. For this choice of parameters two behaviors for weak and strong interactions are approximately consistent with each other (see Figs. \ref{fig:ThrInfW}, \ref{fig:ThrInf}) at the border line between two behaviors set at $\Gamma_{c}=Ue^{-1}$ where the product $\Gamma_{c}\ln(U/\Gamma_{c})$ reaches its maximum with respect to $\Gamma_{c}$. 

The dependencies of the localization threshold $\Gamma_{c}$ on the random field strength $W$ are shown in Fig. \ref{fig:ThrInfW}. As it can be expected the localization domain increases with increasing random field. At small fields, $W < J$, the dependence of localization threshold on the field is weak because disordering is mostly induced by the spin-spin interaction. At larger fields localization threshold shows a universal dependence on the relative disordering parameter $W/N$ increasing proportionally to $W/N$ in the strong interaction case ($W< UN$) and proportionally to $(W/N)^2$ in the weak interaction case ($UN<W$). 

The dependence of the localization threshold on the inter-spin interaction $U$ shown in Fig. \ref{fig:ThrInf} is more complicated. In the weak interaction regime $U<W/N$ the localization threshold $\Gamma_{c}$ decreases with increasing $U$ as $U^{-1}$, emphasizing the dramatic significance of interactions for the delocalization. In the intermediate regime $W/N<U<W/\sqrt{N}$ the localization threshold decreases logarithmically with increasing interaction. Here the interaction strengthens delocalization while suppressing  destructive interference (Fig. \ref{fig:Interf}. b). In these regimes, the localization threshold shows universal behavior at fixed relative disorder strength $W/N$. At strongest interaction $U>W/\sqrt{N}$ localization threshold increases proportionally to the interaction because interaction determines disordering in this regime.

The consideration of the localization transition at infinite temperature is the main goal of the present work. Below I briefly discuss the extensions of theory to finite temperatures and power law interactions.


\section{Finite temperature}
\label{Sec:T}

Many-body localization should be sensitive to the temperature. Indeed, the number of accessible states with close energies decreases with temperature which should make  the localization easier. It is also natural to expect a strong sensitivity of localization to a spin glass transition that can take place in the model under consideration. Below the high temperature paramagnetic phase and low temperature glass phase are discussed separately. 


\subsection{Paramagnetic phase}

The theory can be extended in an almost straightforward manner to the paramagnetic phase of the spin glass model without a longitudinal field, $W=0$. The forward approximation can be applied to that model  using the distribution  of $n$ transition energies $\epsilon_{1,...n}$ near $0$ (cf. Eq. (\ref{eq:SpFlipEDistr1})) evaluated in the Supporting Information, Sec. \ref{sec:SI1} as 
\begin{eqnarray}
P_{n}(0, 0,...)\approx  p_{T}(0)^{n}\eta_{f}(n)^{\frac{nT_{f}^2}{T^2}}. 
\label{eq:SpFlipEDistrT}
\end{eqnarray}
Here $p_{T}(0)$ is the temperature dependent probability of a zero spin flip energy  given by \cite{Sherrington08}
\begin{eqnarray} 
p_{T}(0)=p(0)e^{-\frac{T_{f}^2}{2T^2}},
\label{eq:ZerTrEn}
\end{eqnarray}
and the second factor in the right hand side of Eq. (\ref{eq:SpFlipEDistr1}) accounts for the contribution of spin-spin correlations to the probability distribution. The factor $\eta_{f}(n)$ approaches unity for $n \ll N$. It decreases with increasing $n$ reaching the minimum,  $\eta_{f}(N)=0.7$, at $n\approx N$. 

Using Eq. (\ref{eq:SpFlipEDistrT}) one can estimate the temperature dependent localization threshold  in the $n^{th}$ order forward approximation as (cf. Eq. (\ref{eq:nth_order})) 
\begin{eqnarray}
\Gamma_{cn}(T) \approx \left(\frac{N!}{(N-n)!N^{n}}\right)^{\frac{1}{n}}\frac{\left(e^{\frac{1}{2}}\eta_{f}(n)\right)^{\frac{T_{f}^2}{T^2}}}{4Np(0)\ln\left[\sqrt{nN}\right]}.  
\label{eq:GamnthT}
\end{eqnarray}

Since the forward approximation is not fully applicable to very large $n \sim N$  one can describe the localization threshold similarly to Eq. (\ref{eq:StrIntAns}) introducing  two unknown parameters of order unity 
\begin{eqnarray}
\Gamma_{c}(T) \approx \frac{\eta \left(e^{\frac{1}{2}}\eta_{f}\right)^{\frac{T_{f}^2}{T^2}}}{4Np(0)\ln\left[N\right]},  
\label{eq:GamthT}
\end{eqnarray}
where the factor $\eta$ accounts for the factorial term contribution and the factor $\eta_{f}$ is due to correlations in spin flip energies. Since the latter factor originates from the factor $\eta_{f}(n)$ in Eq. (\ref{eq:GamthT}) chosen at some intermediate $n$ one can expect  $0.7<\eta_{f}<1$. Consequently, for different values of this factor the temperature dependence of the localization threshold in Eq. (\ref{eq:GamthT}) can vary between $1.1^{(T_f/T)^{2}}$ and  $1.65^{(T_f/T)^{2}}$, always leading to the increase of the localization threshold with decreasing the temperature.  In Fig. \ref{fig:OmcT} the temperature dependence of the localization threshold for $W=0$ is shown using the previously chosen parameter $\eta=2$ and the correlation parameter $\eta_{f}=1$ corresponding to the strongest temperature dependence. More accurate analysis of the localization threshold temperature dependence awaits further numerics. 

If the localization transition takes place at some temperature $T$ ($T_{f}<T$) the system possesses the mobility edge at the corresponding energy which can be expressed as $E=-NJ^2/(2k_{B}T)$ \cite{Sherrington75}.   The states in the energy domain $(-E, E)$ are delocalized while other states are localized. 

This observation of many-body mobility edge is   consistent with earlier findings in a random energy model \cite{Laumann14,Laumann16} and conflicts with the general arguments of Ref. \cite{Muller16MobEdge} suggesting that the local
fluctuations into the ergodic phase within the supposedly localized phase can serve as mobile bubbles that induce global delocalization. Such local fluctuations are lacking in the present model possessing an infinite interaction radius. The extension of the mobility edge consideration to the systems with short-range interactions is beyond the scope of the present work. 

In the case of strong random fields $W \gg J$, yet strong interactions, $W<UN$ (Eq. (\ref{eq:n2StrInt0}), the system undergoes a spin glass transition  at the temperature $T_{f} \approx 4J^2/(3\sqrt{2\pi}Wk_{B})$ (cf. Ref. \cite{Hadjiagapiou14}). In a paramagnetic phase one can expect the increase of the localization threshold with decreasing the temperature qualitatively similar to Eq. (\ref{eq:GamthT}) due to the reduction of the phase space.  The accurate analysis of this dependence is beyond the scope of the present work.
For the sake of simplicity the temperature dependencies in paramagnetic phases for finite random fields are skipped in Fig. \ref{fig:OmcT}. 

If the interaction is weak there is no spin glass transition and spin-spin correlations can be approximately neglected. Then the infinite temperature results, Eq. (\ref{eq:WeakIntAns}), should be approximately valid until thermal energy is much bigger that the threshold field $\Gamma_{c}$ ($\Gamma_{c} \approx W/(JN^{3/2}) \ll k_{B}T$). At lower temperature the localization should take place at $\Gamma < k_{B}T$ while at $\Gamma >k_{B}T$ spins are frozen out in their ground states. Therefore I expect the system to be localized   at any transverse field in the low temperature limit $k_{B}T \ll W/(JN^{3/2})$.

\subsection{Spin glass phase}

In a spin glass phase the system occupies one of the local energy minima (valley), where most of spins are frozen out \cite{Parisi80,Sherrington75}. Yet there are spins with energies comparable to the thermal energy which can be in both localized or delocalized states. The number of such ``thermal" spins $N_{T}$ can be approximately evaluated using the spin flip  probability energy density $p_{T}(0)$ at zero energy as $N_{T}=Np_{T}(0)k_{B}T$. This probability density  has been determined numerically in the spin glass phase (in the absence of longitudinal field) as \cite{Sherrington08}  (see also Ref.  \cite{Parisi80}) 
\begin{eqnarray}
p_{T}(0)=p_{f}(0)T/T_{f},
\label{eq:SGDoS}
\end{eqnarray}
where $p_{f}(0)$ refers to the glass transition point $T=T_{f}$. The interaction of spins $J_{T}=J\sqrt{N_{T}/N}$ scales as the thermal energy suggesting that these ``thermal" spins are near the spin glass transition in the subsystem limited to those spins \cite{Parisi80}. Consequently the localization transition for these spins can be approximately described using Eq. (\ref{eq:GamnthT}) with the factor $p_{T}(0)$ given by Eq. (\ref{eq:SGDoS}). This yields 
\begin{eqnarray}
\Gamma_{c} \approx \frac{\eta \left(\eta_{f}\right)^{\frac{T_{f}^2}{T^2}}}{4Np_{T}(0)\ln\left[N_{T}\right]},  
\label{eq:GamthTSG}
\end{eqnarray} 
Since the spin glass phase is universal  with respect to random longitudinal fields \cite{Hadjiagapiou14} the same expression should be approximately valid for spin glasses subjected to random longitudinal fields so many-body localization transitions at different field strengths $W$ should be described by the same equation below the glass transition temperature.

\begin{figure}[h!]
\centering
\includegraphics[width=\columnwidth]{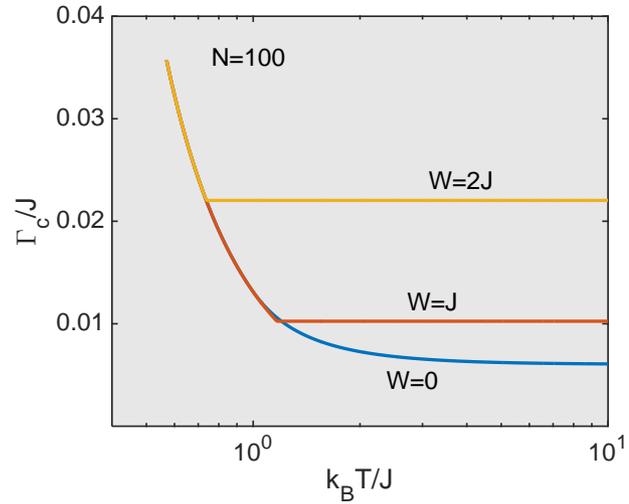}
\caption{\small The localization threshold vs. temperature in the regime of strong interaction for $N=100$ spins and different strengths of random field (Eqs. (\ref{eq:GamthT}), (\ref{eq:GamthTSG})).}
\label{fig:OmcT}
\end{figure}

The temperature dependence of the localization threshold is shown in Fig. \ref{fig:OmcT} for different strengths of random fields ($\eta=2$ and $\eta_{f}=1$ are chosen as previously discussed). The results are applicable until the interspin interaction energy is less than the thermal energy $J/\sqrt{N} \ll k_{B}T$. At lower  temperatures the system occupies one of the degenerate ground states. The delocalization described above includes only around $2^{N_{T}}$ many-body states corresponding to the local energy minimum under consideration while other minima are separated from it by macroscopic energy barriers. Since the statistics  of overlaps between different valleys and potential barriers separating them is not quite clear \cite{Parisi80,Moore89SGBarriers,Young82SGOverStat,Young83BarrVall} the analysis of the problem of coupling between different valleys due to the transverse field in the spin glass phase is left for future studies.

\section{Power law interactions}
\label{sec:PowLaw}

Here the case of the power law interaction between spins is considered qualitatively in the model Eq. (\ref{eq:H}) with a  modified distance dependent interspin interaction,
\begin{eqnarray}
J_{ij}=j_{ij}(a/r_{ij})^{\alpha}, 
\label{eq:PowLaw}
\end{eqnarray}
in a d-dimensional lattice with the period $a$. Interaction constants $j_{ij}$ are set to be all random, uncorrelated parameters of order of $J_{0}$. A random field is assumed to be strong,   $W \gg J_{0}$. It is also assumed that $\alpha>d$ to avoid a single particle delocalization \cite{Anderson58,Levitov90}. 

Consider the effect of interaction at long distances on a delocalization. For the sake simplicity one can replace all interactions with the one at the maximum distance $r \sim aN^{\frac{1}{d}}$ ($J_{ij} \sim J_{0}N^{-\frac{\alpha}{d}}$). 

It turns out that the modified system matches the weak interaction limit of the spin glass model with the long range interactions since the interspin interaction is less than $W/N$ (remember that it is assumed that $W>J_{0}$ and $\alpha > d$). Then one can estimate the localization threshold using Eq. (\ref{eq:WeakIntAns}) as
\begin{eqnarray}
\Gamma_{c}\sim \frac{W^2}{N^{\frac{2d-\alpha}{d}} J_{0}}.         
\label{eq:AnsPowLawInt1}
\end{eqnarray}

Eq.  (\ref{eq:AnsPowLawInt1}) predicts the localization threshold decreasing to zero in the thermodynamic limit ($N\rightarrow \infty$) if the interaction exponent is less than twice the system dimension ($\alpha<2d$) in a full accord with the early theory's predictions \cite{ab98book}, while in the opposite case the theory is not applicable since $\Gamma_{c}$ exceeds the random field strength $W$. In the case $\alpha<2d$ the threshold  field decreases with the system size as $N^{-\frac{2d-\alpha}{d}}$ in a full  agreement with Ref. \cite{ab06preprint}. Also it turns out that the parametric dependence of the estimate Eq. (\ref{eq:AnsPowLawInt1}) is consistent with the analysis of Ref. \cite{ab15MBL}. 
This can be shown by considering only resonant spins ($|\epsilon_{i}|<\Gamma$, see Eq. (\ref{eq:SpFlipE})) and introducing the flip-flop interaction between them following Ref. \cite{ab06preprint}. The number of these spins s given by $N_{*} \sim N\Gamma p(0)$, their interaction at the average distance takes the form  $\tilde{U} \sim J_{0} (p(0)\Gamma)^{\frac{\alpha}{d}}$ and their characteristic random energy is already chosen to be $W_{*} \sim \Gamma$. Applying the delocalization criterion of Ref. \cite{ab15MBL} ($W_{*} \sim \tilde{U}N^{\frac{2d-\alpha}{d}}$, see the last column in the Table 1 there) one reproduces  Eq. (\ref{eq:AnsPowLawInt1}). Thus the suggested method leads to the results  agreeing with the previous studies so it is potentially extendable to short-range interactions. 


\section{Conclusions}
\label{sec:Concl}

The localization-delocalization transition is investigated in a Sherrington-Kirkpatrick spin glass model with  random longitudinal and small transverse fields. The localization threshold expressed in terms of the critical transverse  field $\Gamma_{c}$  has been estimated by exploiting the similarity of the problem to the exactly solvable localization problem on a Bethe lattice. The localization transition is sensitive to the relationship of  inter-spin interactions $U$ and random field $W$, leading to  three distinguishable regimes (see Fig. \ref{fig:ThrInf}). If the typical spin-spin interaction is weak compared to the characteristic minimum random field per spin (weak interaction case $U<W/N$, Sec. \ref{subsec:weak}) the interaction essentially determines delocalization and the critical transverse  field $\Gamma_{c}$ decreases proportionally to the inverse interaction. In the intermediate regime of strong interaction and random fields ($W/N < U < W/\sqrt{N}$) the interaction is responsible for suppression of destructive interference, and the localization threshold depends on it only logarithmically. At strong interaction and weak random field ($W/\sqrt{N} < U$) the effective random field is determined by the interaction, and the localization threshold increases proportionally to the interaction. The localization threshold increases with decreasing temperature, approaching the universal behavior in the spin glass phase (Fig. \ref{fig:OmcT}). 

The technique developed in the present work can be extended to systems with short-range interactions as demonstrated by considering the power law interaction between spins. If the interaction radius $R_{0}$ in a $d$-dimensional system is finite and exceeds an interatomic distance $a$ it is natural to expect that a many body localization transition can be described using the present theory with a modified connectivity parameter $K$ for the matching Bethe lattice problem, i. e. $K=(R_{0}/a)^d$ instead of $K=N$ as in the case of an infinite interaction radius. The modified dependencies should remain qualitatively valid after this substitution. The accurate analysis of this prediction is for future numerical and analytical challenges.  

The results of the present work can be verified experimentally  using ultracold atoms \cite{Lukin14MBLGen,Monro16,LukinDiamond16} and numerically by exact diagonalization. 


This work is partially supported by the National Science Foundation (CHE-1462075) and the Tulane University Carol Lavin Bernick Faculty Grant. Author also acknowledges Max Planck Institute for Physics of Complex Systems Visitor Program and Karlsruhe Institute of Technology for supporting his visits and Alexander Mirlin, Giuseppe  Detomasi, Noah Rahman, Kelly Dougherty, Frank Pollmann, Antonello Scardicchio, Christopher Laumann,  Achilleas Lazarides, Igor Gornyi, Ivan Protopopov, Dmitry Polyakov and Peter Fulde for very useful discussions and critical remarks. 

\bibliographystyle{andp2012}
\bibliography{MBL}

\section{Supplementary Materials}

\subsection{Correlations of spin-flip energies}
\label{sec:SI1}

Here the probability distribution of spin-flip transition energies is considered in a spin glass model without random fields at infinite and finite temperatures $T$, yet exceeding the spin glass transition temperature $T_{f}=J/k_{B}$. It is shown that the correlations between energies can be always neglected at an infinite temperature. At finite temperatures $T>T_{f}$ they can be also neglected if the number of spin flips $n$ is less than the total number of spins $N$ while for $n \sim N$ they can slightly reduce the localization threshold. In the infinite temperature case the generalization of the results to arbitrarily external fields is straightforward, while the consideration of  the  finite temperature case in the presence of an external field is beyond the scope of the present work. 

Consider $n$ subsequent spin flips of a spin sequence $\{\sigma_{p}\}$, $p=1,...n$, in some initial Ising product state $a$ corresponding to the temperature $T$, $T>T_{f}$ characterized by the spin projections $\{\sigma_{ia}\}$ ($i=1,...n$). Then the energy change $E_{p}$ after flips of first $p$ spins can be expressed as 
\begin{eqnarray}
E_{p} = 2\sum_{l=p+1}^{N}\sigma_{la} \sum_{k=1}^{p}J_{kl}\sigma_{ka}.  
\label{eq:EpTSI}
\end{eqnarray}
The first sum is taken over all $N-p$ remaining spins. To evaluate the joint distribution function, $P_{n}$ of all flip energies $E_{p}$  entering the forward approximation in the case of interest where these energies are close to zero one can use its Fourier transform as 
\begin{eqnarray}
P_{n}=\frac{1}{(2\pi)^n}\prod_{m=1}^{n} \int_{-\infty}^{\infty}dt_{m}F(t_{1},..t_{n}), 
\nonumber\\
F(t_{1},..t_{n})
=\left<e^{i\sum_{p=1}^{n}t_{p}E_{p}}\right>,   
\label{eq:EpTFTSI}
\end{eqnarray}
where the averaging is always defined as $<...> = {\rm Tr }\left[ e^{-\beta H_{0}}...\right]$ and $\beta=1/(k_{B}T)$ while the contribution of the transverse field $\Gamma \sim \Gamma_{c}$ to thermal averages can be neglected since $\Gamma_{c} \ll k_{B}T$. This result should be compared with the distribution assuming independent probabilities for all $n$ energy changes used in the main body of the manuscript which suggests (this distribution is expressed in the main text in terms of the parameter  $W_{1}=2J$ in the case $W=0$)
\begin{eqnarray}
P_{n0}=p_{T}(0)^{n}=\frac{1}{(8\pi)^{\frac{n}{2}}J^{n}}e^{-n\frac{J^2}{2(k_{B}T)^2}}.   
\label{eq:EpTIndSI}
\end{eqnarray}

Since at $T>T_{g}$ spin flip energies obey Gaussian statistics \cite{Sherrington75,Parisi80} one can express the Fourier transform in Eq. (\ref{eq:EpTFTSI}) as 
\begin{eqnarray}
F(t_{1},..t_{n})=e^{i\sum_{p=1}^{n}t_{p}<E_{p}>-\frac{1}{2}\sum_{p,q}^{n}t_{p}t_{q}<\delta E_{p}\delta E_{q}>}, 
\nonumber\\
\delta E_{p}=E_{p}-<E_{p}>.   
\label{eq:EpTFT2SI}
\end{eqnarray}
Applying the mean field approximation to spin averages \cite{Sherrington08} and replacing the parameters $J_{ij}^2$ with their averages $J^2/N$ (this replacement is justified by the law of large numbers for $N, n \gg 1$ and it can still be used as an estimate in the case of $n \sim 1$) one can express the exponent in the Fourier transform as 
\begin{eqnarray}
F(t_{1},..t_{n})=e^{2i\frac{J^2}{k_{B}T}\sum_{p=1}^{n}t_{p}v_{p}-\frac{4J^2}{2}\sum_{p,q}^{n}t_{p}M_{pq}t_{q}}, 
\nonumber\\
v_{p}=p\left(1-\frac{p}{N}\right), ~ M_{pq}={\rm min}(p, q)\left(1-\frac{{\rm max} (p, q)}{N}\right).  
\label{eq:EpTFT3SI}
\end{eqnarray}
One can substitute Eq. (\ref{eq:EpTFT3SI}) into Eq. (\ref{eq:EpTFTSI}) and perform integration over all variables $t_{p}$. This yields 
\begin{eqnarray}
P_{n}=\frac{\exp\left[-\frac{J^2}{2(k_{B}T)^2}\sum_{p,q}^{n}\left[M^{-1}\right]_{pq}v_{p}v_{q}\right]}{(8\pi)^{\frac{n}{2}}J^{n}\sqrt{{\rm det}(\widehat{M})}}.   
\label{eq:EpTFAnsSI}
\end{eqnarray}

The localization threshold behavior  can be described within the $n^{th}$ order forward approximation as $\Gamma_{cn} \propto 1/(P_{n})^{\frac{1}{n}}$ (see Sec. \ref{sec:SGinfT} in the main text). Then the correlations modify the localization threshold by the factor 
\begin{eqnarray}
\eta(n) = \left(\frac{P_{n0}}{P_{n}}\right)^{\frac{1}{n}} 
\nonumber\\
= \left({\rm det}(\widehat{M})\right)^{\frac{1}{2n}}\exp\left[\frac{J^2}{2T^2}\left(\frac{1}{n}\sum_{p,q}^{n}\left[M^{-1}\right]_{pq}v_{p}v_{q}-1\right)\right]. 
\label{eq:EpTFAns2SI}
\end{eqnarray}

If one neglects $p/N \ll 1$ and $q/N \ll 1$ in the definitions of the vector $v$ and matrix $\widehat{M}$ in Eq. (\ref{eq:EpTFT3SI}) then it can be shown  that $\eta=1$ (see the main body of the manuscript) and there are no corrections to the localization threshold due to correlations. Consequently the correlation effect can be significant only for the number of spin flips comparable to the total number of spins, $n \sim N$. Moreover since for the corrected matrix $\widehat{M}$ one has ${\rm det}(\widehat{M})=1-n/N$ the correction from the  factor containing the determinant, $(1-n/N)^{\frac{1}{2n}}$, is always close to $1$ at sufficiently large $N$ except for the case $n=N$ so the factor containing the determinant can be skipped. Since there are no other corrections at an infinite temperature, the correlations in spin-flip energies cannot affect the localization threshold in that case. 

\begin{figure}[h!]
\centering
\includegraphics[width=\columnwidth]{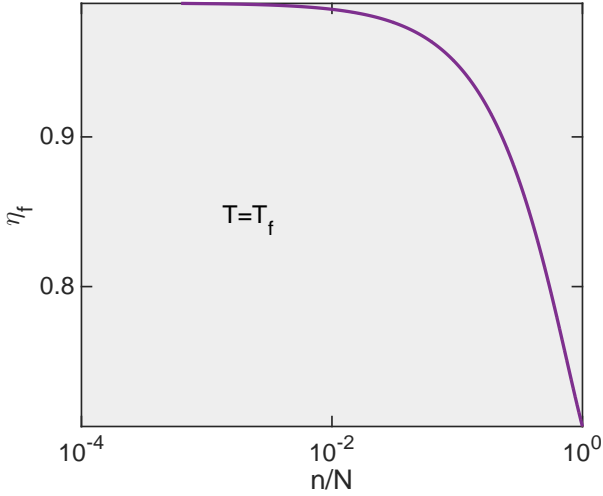}
\caption{\small The correction factor $\eta_{f}$ calculated numerically using the exponential factor in  Eq. (\ref{eq:EpTFAns2SI}) versus а relative number of spin flips, $n/N$, calculated for the numbers of spins $N=100, 400, 900, 1600$ at the transition temperature $T=T_{f}$. All data belongs to an almost universal curve. In the case of $n \ll N$ this factor rapidly approaches unity.}
\label{fig:CorrTemp}
\end{figure}

The temperature dependent exponent in the right hand side of Eq. (\ref{eq:EpTFAns2SI})  differs from $1$ at $n \sim N$ as shown in Fig. \ref{fig:CorrTemp} where the factor $\eta$ is given at the glass transition temperature $T=T_{f}=J/k_{B}$ (it is denoted as $\eta_{f}$ there and in the main text). The temperature dependent  factor $\eta(n)$  in Eq. (\ref{eq:EpTFAns2SI}) can be expressed for $T>T_{f}$ as $\eta_{f}(n)^{\frac{T_{f}^2}{T^2}}$ and this equation is used in the main body of the present work. 

At small number of flips, $n \ll N$, one has  $\eta_{f}(n) \approx \exp (-n/(2N)) \sim 1$  while this factor approaches $0.7$ at $n \rightarrow N$.  This behavior can lead to some reduction of the localization threshold at a finite temperature as discussed in the main text.

\subsection{Localization in the Bethe lattice with correlated site energies.}
\label{sec:BLSol}

In contrast with the standard Bethe lattice problem \cite{AbouChacra73,Laumann14} with all non-correlated site energies $\Phi_{a}$ the energies of adjacent  sites of the spin glass model 
\begin{eqnarray}
\widehat{H}=\widehat{H}_{0}+\widehat{V},
\nonumber\\
\widehat{H}_{0}=\sum_{i=1}^{N}\phi_{i}\sigma_{i}^{z} + \sum_{i=1}^{N}\sum_{j=1}^{i-1}J_{ij}\sigma_{i}^{z}\sigma_{j}^{z}, ~ \widehat{V}=\Gamma\sum_{i}\sigma_{i}^{x}. \label{eq:HSI}
\end{eqnarray}
given by eigenvalues of the Ising Hamiltonian $\widehat{H}_{0}$ different by a single spin flip are strongly correlated. Indeed the change of the Ising energy $\epsilon_{i}$ defined as 
\begin{eqnarray}
\epsilon_{i}=2\sigma_{ia}(\phi_{i}+\sum_{k}J_{ik}\sigma_{ka})
\label{eq:SpFlipESI}
\end{eqnarray}
due to a flip of some spin $i$ is of order of $W_{1}=2\sqrt{W^2+J^2}$ which is much less than  the typical energy of the state $a$ ($\Phi_{a}\sim W_{1}\sqrt{N} \gg W_{1}$ at $T=\infty$).  Consequently the accurate solution of Ref.  \cite{Laumann14} is not applicable to the model Eq. (\ref{eq:HSI}). 

However,  the accurate solution is still possible if the correlations between different transition energies $\epsilon_{i}$ Eq. (\ref{eq:SpFlipESI}) can be neglected. This is indeed the case for the large coordination number $N\gg 1$. In the absence of spin-spin correlations (an infinite temperature) the average squared transition energy is given by $<\epsilon_{j}^2>=W_{1}^2$ while the correlation energy for two transition energies associated with flips of spins $i$ and $j$ is  given by $<\epsilon_{i}\epsilon_{j}>=4J^2/N \ll J^2 \ll W_{1}^2$.  Then following the law of large numbers one can describe the statistics of transition energies  $\epsilon$ in Eq. (\ref{eq:SpFlipESI}) using independent Gaussian distributions 
\begin{eqnarray}
p(\epsilon)=\frac{1}{\sqrt{2\pi}W_{1}}e^{-\frac{\epsilon^2}{2W_{1}^2}}, ~ W_{1}=2\sqrt{W^2+J^2}. 
\label{eq:SpFlipEDistrSI}
\end{eqnarray}
(see also the consideration in the previous section). 

Consider the model of fully uncorrelated transition energies independent of site energies. Then one can fix the energy of an arbitrarily Bethe lattice  site at $\Phi=0$ and express the energy of any other site separated from the given site by $n$ steps as a sum of $n$ random uncorrelated transition energies. In the limit of an infinite Bethe lattice all possible energies in the domain $(-\infty, \infty)$ have equal chances to be realized so the site energy density of states $P(E)$ approaches zero which corresponds to the infinitely strong disordering. In the case of uncorrelated site energies this should lead to full localization \cite{Laumann14}. However, this is not the case for correlated site energies. 
 
Following the self-consistent theory of localization \cite{AbouChacra73} one can express the single site $a$ diagonal Green function as $G_{a}=(E-\Phi_{a}-\Sigma_{a})^{-1}$ where $\Sigma_{a}$ is the self-energy. 
The self-consistent equation linearized with respect to the imaginary parts of self-energies vanishing below the localization transition can be then written as \cite{AbouChacra73} (more explanation is provided in the next section, where the forward approximation is considered)
\begin{eqnarray}
{\rm Im}~\Sigma_{a}= \Gamma^2\sum_{k}\frac{{\rm Im} ~\Sigma_{a_{k}}}{(E-\Phi_{a}-E_{k}-{\rm Re}~\Sigma_{ak})^2}. 
\label{eq:AbouChacra}
\end{eqnarray}
Here the difference in coordination numbers between $N$ and $N-1$ is ignored since $N \gg 1$ (cf. Ref. \cite{AbouChacra73}). The delocalization transition takes place at $\Gamma=\Gamma_{c}$ where the linearized equation for imaginary parts (Eq. (\ref{eq:AbouChacra}) acquires the first non-zero solution.

Using the ansatz of Ref. \cite{AbouChacra73} $\left<e^{-s{Im\Sigma_{a}}}\right>=1-s^{\beta}A(R_{a})$ ($R_{a}=E-\Phi_{a}$) valid for $s\rightarrow 0$ and neglecting ${\rm Re} \Sigma_{ak}$ terms in the denominator of the right hand side of Eq. (\ref{eq:AbouChacra}) one can get the approximate equation for the function $A$ in the form 
\begin{eqnarray}
A(R)=N\Gamma^{2\beta}\int_{D} \frac{p(\epsilon)A(R-\epsilon)d\epsilon}{|R-\epsilon|^{2\beta}},  
\label{eq:AbouChacra2}
\end{eqnarray}
where $\epsilon$ stands for spin flip transition energies within the Bethe lattice and $p(\epsilon)$ is the distribution function of those energies Eq. (\ref{eq:SpFlipEDistrSI}). The exponent $\beta$ should be optimized to maximize the threshold field $\Gamma_{c}$ corresponding to the first non-zero solution of Eq. (\ref{eq:AbouChacra}). The integration domain $D$ is chosen as $(-\infty, R-p(0)\Gamma^2) \cup (R+p(0)\Gamma^2, \infty)$ in accordance with Refs. \cite{Anderson58,AbouChacra73} to avoid the anomalously large contributions $\Gamma^2/|R-\epsilon|>1/p(0)$ to a real part of a self-energy. This constraint reflects level repulsion restricting the minimum energy difference of coupled states. 

Under these conditions the real part of the self-energy can be neglected. This simplification is justified within logarithmic accuracy by a large coordination number    $N \gg 1$ where the real part of the self-energy ($Np(0)\Gamma_{c}^2$) is much smaller than the typical energy ($1/p(0)$). The latter condition is satisfied  near the localization transition point $\Gamma_{c} \sim 1/(Np(0))$ for $N \gg 1$. 

In the case of $N \gg 1$ one has $1-2\beta \ll 1$ \cite{AbouChacra73} so the integral in Eq. (\ref{eq:AbouChacra2}) over energies is nearly logarithmic. Then setting $R=0$ one can rewrite Eq. (\ref{eq:AbouChacra2}) in the form 
\begin{eqnarray}
A(0)=N\Gamma^{2\beta}\int_{D} \frac{p(\epsilon)A(0)d\epsilon}{|\epsilon|^{2\beta}}
\nonumber\\
+N\Gamma^{2\beta}\int_{D} \frac{p(\epsilon)(A(-\epsilon)-A(0))d\epsilon}{|\epsilon|^{2\beta}}. 
\label{eq:AbouChacra3}
\end{eqnarray}
The second term can be neglected within logarithmic accuracy compared to the first one because of the compensation of the small denominator at $|\epsilon| \rightarrow 0$. Evaluating the remaining integral 
with the same accuracy one can obtain the equation for the localization threshold in the form 
\begin{eqnarray}
1=2Np(0)\Gamma_{c}\frac{(p(0)\Gamma_{c})^{2\beta-1}-(p(0)\Gamma_{c})^{-2\beta+1}}{(1-2\beta)}. 
\label{eq:AbouChacra4}
\end{eqnarray}
Similarly to Ref. \cite{AbouChacra73} (see Sec. 6 there) one can show that the minimum of the right hand side of Eq. (\ref{eq:AbouChacra4}) is 
realized at $\beta = 1/2$ leading to the familiar definition of the localization transition \cite{AbouChacra73,Laumann14}
\begin{eqnarray}
\Gamma_{c1}=\frac{1}{4Np(0)\ln(1/(p(0)\Gamma_{c}))} \approx \frac{1}{4Np(0)\ln(N)},  
\label{eq:AbouChacra41}
\end{eqnarray}
where the typical random energy $1/p(0)$ is given by the characteristic spin flip  energy rather than the site energy which is macroscopically large \cite{Laumann14}. 

The difference $e/2$ between the prefactors in the result of Ref. \cite{Laumann14} and in Eq. (\ref{eq:AbouChacra41})  has the same origin as in two different estimates in Ref. \cite{AbouChacra73} where the integration constraint is included (Sec. 6 there, cf. Eq. (\ref{eq:AbouChacra41})) or ignored (Sec. 5 there). The integration constraint   is included in the present work since it gives a better estimate for the localization transition in the Bethe lattice \cite{AbouChacra73}. The addition of a similar constraint to the analysis of Ref. \cite{Laumann14} will change the estimate for the localization threshold there by the same factor so two approximations are technically equivalent. 

Correlations in site energies result in a dramatic suppression of localization in agreement with previous studies (see e. g. the review \cite{Izrailev12} and references therein). Further applications of the proposed solution to the localization problem with correlated disorder can be of interest. 

\subsection{Localization threshold in the case of a strong interaction}
\label{sec:SIStrong}

\subsubsection{Self-consistent forward approximation for a spin glass problem}

The derivation begins with the equation for the Green functions in the basis of Ising states (spin projection sequences $\{a\}$) defined as $G_{ab}=\left<a\left|(E-\widehat{H})^{-1}\right|b\right>$, where $E$ is the energy of the state of interest and the Hamiltonian $\widehat{H}$ is defined by Eq. (\ref{eq:HSI}). 
The diagonal Green function ($G_{a}=G_{aa}$) satisfies the equation 
\begin{eqnarray}
(E-\Phi_{a})G_{a}=1+\Gamma\sum_{k}G_{a_{k}a}, 
\label{eq:GFEq1}
\end{eqnarray}
where $\Phi_{a}=\sum_{i}\phi_{i}\sigma_{ia}+(1/2)\sum_{ij}J_{ij}\sigma_{ia}\sigma_{ja}$ is the site energy expressed in terms of the spin projections to the z-axis ($\sigma_{ia}$) for the specific state $a$. Each state $a_{k}$ (neighboring site in the Bethe lattice) can be obtained from the state $a$ by flipping the spin $k$. Green functions in the right hand side of Eq. (\ref{eq:GFEq1}) obey the equations  
\begin{eqnarray}
(E-\Phi_{a,k})G_{a_{k}a}=\Gamma G_{a} +\Gamma\sum_{k'\neq k}G_{a_{kk'}a}, 
\label{eq:GFEq2}
\end{eqnarray}
 where the states $a_{kk'}$ are obtained from the state $a$ by flipping spins $k$ and $k'$. The self-consistent equation \cite{AbouChacra73} can be derived for the Bethe lattice problem from Eq. (\ref{eq:GFEq1}) introducing the diagonal Green function $G_{k}$ in the state $a_{k}$ (for the sake of simplicity $G_{a_{k}a_{k}}$ is denoted as $G_{k}$) ignoring the connection of this state to the state $a$ which permits one to express the function $G_{a_{k}a}$ as 
 \begin{eqnarray}
G_{a_{k}a}=\Gamma G_{a} G_{k}.  
\label{eq:GFEq3}
\end{eqnarray}
If the function $G_{a} $ is taken by ignoring $1$ out of $N$  connections in the related graph  (this corrresponds to the coordination number $N-1$) the substitution of Eq. (\ref{eq:GFEq3}) into Eq. (\ref{eq:GFEq2}) leads to the self-consistent equation of Ref. \cite{AbouChacra73}. In our case of $N\gg 1$ the difference of a single connection can be neglected and one can express Eq. (\ref{eq:GFEq3}) in the self-consistent form  \cite{AbouChacra73} 
 \begin{eqnarray}
G_{a}=\frac{1}{E-\Phi_{a}-\Gamma^2\sum_{k}G_{k}}.
\label{eq:GFEq4}
\end{eqnarray}
The analysis of this equation performed in the previous section leads to the localization threshold given by Eq. (\ref{eq:AbouChacra41}) for the matching Bethe lattice problem. To examine the validity of the self-consistent approximation one can make one or more iterative steps expanding the off-diagonal Green functions $G_{a_{k}a}$ in the same manner as it was done in Eq. (\ref{eq:GFEq3}). As  shown below these iterations lead to the forward approximation.

\begin{figure}[h!]
\centering
\includegraphics[width=\columnwidth]{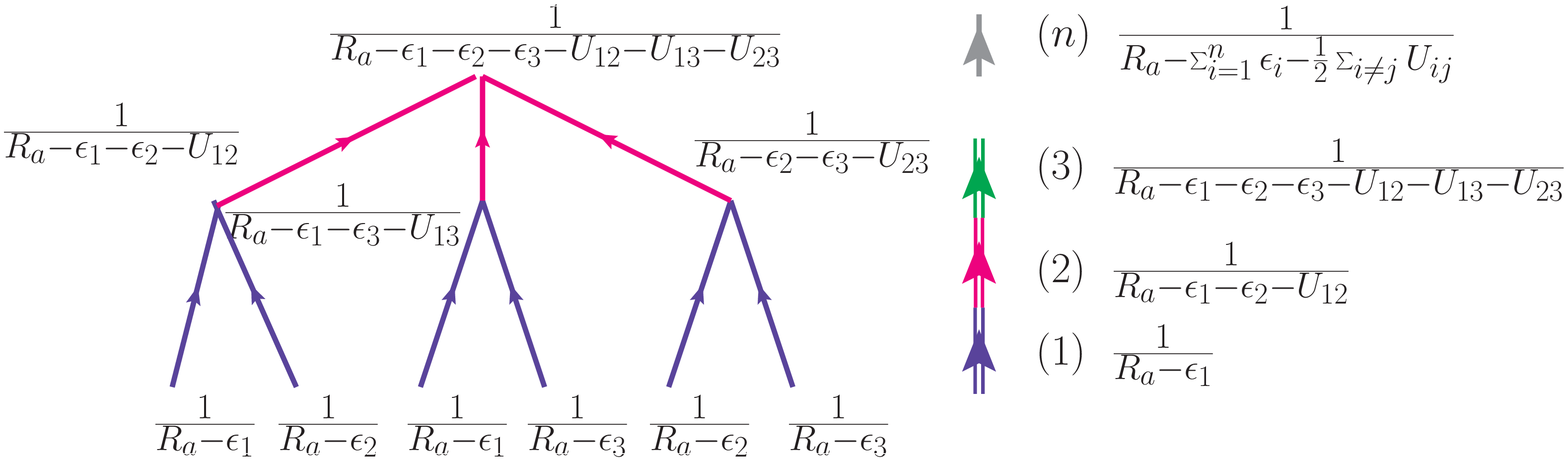}
\caption{\small The third order forward approximation and  a general structure of the $n^{th}$ order term (Eq. (\ref{eq:AbouChacraModnSI})) determined by sums of products of fractions taken along the paths from the top to the bottom, $R_{a}=E-\Phi_{a}$.}
\label{fig:nord}
\end{figure}
 
 Consider the second iteration in detail. It can be performed using the equation for the off-diagonal Green function $G_{a_{kk'}a}$ which can be written as 
 \begin{eqnarray}
\frac{E-\Phi_{a_{kk'}}}{\Gamma}G_{a_{kk'}a}= G_{a_{k}a}+G_{a_{k'}a}+
\sum_{k''\neq k,k'}G_{a_{kk'k''}a}. 
\label{eq:GFEq5}
\end{eqnarray}
Performing the same trick as in Eq. (\ref{eq:GFEq3}) one can express the solution to Eq. (\ref{eq:GFEq5}) as 
 \begin{eqnarray}
 G_{a_{kk'}a}=\Gamma (G_{a_{k}a}+G_{a_{k'}a}) G_{kk'}, ~  G_{kk'}=G_{a_{kk'}a_{kk'}}.
\label{eq:GFEq7}
\end{eqnarray}

Introducing the function $\lambda_{k}=G_{a_{k}a}/(\Gamma G_{a})$ and substituting Eq. (\ref{eq:GFEq7}) into Eq. (\ref{eq:GFEq2}) one recovers the form 
\begin{eqnarray}
(E-\Phi_{a_k})\lambda_{k}=1 +\Gamma^2 \sum_{k'\neq k}G_{kk'}(\lambda_{k}+\lambda_{k'}), 
\label{eq:GFEq8}
\end{eqnarray} 
while the initial diagonal  Green function can be expressed as 
$G_{a}=(E-\Phi_{a}-\Gamma^2\sum_{k}\lambda_{k})^{-1}$.

 The localization transition is determined by the imaginary part of the self-energy 
 ${\rm Im} \Sigma_{a}=\Gamma^2 \sum_{k}{\rm Im}\lambda_{k}$, 
 while the real and imaginary parts of functions $\lambda_{k}$ satisfy equations following from Eq. (\ref{eq:GFEq8}), which can be written as 
 \begin{eqnarray}
 {\rm Re}~\lambda_{k}=\frac{1}{E-\Phi_{a_k}} 
 +\frac{1}{E-\Phi_{a_k}}\Gamma^2 \sum_{k'\neq k}{\rm Re} G_{kk'}{\rm Re} (\lambda_{k}+\lambda_{k'}), 
 \nonumber\\ 
 {\rm Im}~\lambda_{k}=\frac{1}{E-\Phi_{a_k}} \Gamma^2 \sum_{k'\neq k}{\rm Im} G_{kk'}{\rm Re} (\lambda_{k}+\lambda_{k'}) 
 \nonumber\\
  +\frac{1}{E-\Phi_{a_k}} \Gamma^2 \sum_{k'\neq k}{\rm Re} G_{kk'}{\rm Im} (\lambda_{k}+\lambda_{k'}).
\label{eq:GFEq8a}
\end{eqnarray} 

The forward approximation can be derived from this equation as follows. First the real part of the function $\lambda$ is calculated to lowest order in $\Gamma$ ignoring the real part of the self-energy similar to Ref. \cite{AbouChacra73} where it is justified in the case of a large coordination number (here the coordination number is $N\gg 1$). Then the first line in Eq. (\ref{eq:GFEq8a}) takes the form 
 \begin{eqnarray}
 {\rm Re}~\lambda_{k}=\frac{1}{E-\Phi_{a_k}}.
\label{eq:GFEq9SI}
\end{eqnarray} 
Second the imaginary parts of functions $\lambda_{k}$ corresponding to ``backwards" processes involving several flips of the same spin are neglected on the right hand side of the second line of Eq. (\ref{eq:GFEq8a}). Then using Eq. (\ref{eq:GFEq9SI}) one can represent this second line as 
 \begin{eqnarray}
 {\rm Im}~\lambda_{k}=\frac{1}{E-\Phi_{a_k}} \Gamma^2 \sum_{k'\neq k}{\rm In} G_{kk'} \left(\frac{1}{E-\Phi_{a_k}}+\frac{1}{E-\Phi_{a_k'}}\right).
\label{eq:GFEq10a}
\end{eqnarray}

Substituting the results for ${\rm Im}$ $\lambda_{k} $ into Eq. (\ref{eq:GFEq8}) one can get the modified self-consistent equation for the self-energy in the form 
 \begin{eqnarray}
{\rm Im} ~\Sigma_{a} = \frac{\Gamma^4}{2}\sum_{k,k'}\left(\frac{1}{E-\Phi_{a_{k}}}+\frac{1}{E-\Phi_{a_{k'}}}\right)^2
\frac{{\rm Im}~ \Sigma_{kk'}}{(E-\Phi_{a_{kk'}})^2}, 
\label{eq:AbouChacraMod1SI}
\end{eqnarray}
where $\Sigma_{kk'}$ is the self-energy for the diagonal Green function in the state $a_{kk'}$. The energy $\Phi_{a_{kk'}}$ can be expressed as 
 \begin{eqnarray}
\Phi_{a_{kk'}}=\Phi_{a}+\epsilon_{k}+\epsilon_{k'}-U_{kk'}, ~ U_{kk'}=4J_{kk'}\sigma_{ka}\sigma_{k'a}, 
\label{eq:TransEn2SI}
\end{eqnarray} 
 where $\sigma_{ia}$ is the $z$ axis projection of the spin $i$ in the state $a$.
 
 Similarly one can derive the self-consistent equation performing $n>2$ iterations and ignoring the corrections to the real part of the self-energy and backwards processes. It is convenient to introduce a more general function 
$\lambda_{\{k\}}=G_{a_{\{k\}}a}/(\Gamma G_{a})$ where $\{k\}$ denotes the sequence of spins flipped in the state $a$ forming the state $a_{\{k\}}$. Using this notation one can still express the diagonal Green function as $G_{a}=(E-\Phi_{a}-\Gamma^2\sum_{k}\lambda_{k})^{-1}$.  The function $\lambda_{\{k\}_{p}}$ for  the sequence $\{k\}_{p}$ of $p$ flipped spins ($1<p<n$) obeys the equation ($\Phi_{\{k\}_{p}}$ is a simplified notation for the diagonal energy of the state obtained from the initial state $a$ by means of flipping $p$ spins, belonging to the sequence $\{k\}_{p}$)
 \begin{eqnarray}
(E-\Phi_{\{k\}_{p}})\lambda_{\{k\}_{p}}=\Gamma \sum \lambda_{\{k^{-}\}_{p-1}} +\Gamma \sum \lambda_{\{k^{+}\}_{p+1}},
\label{eq:lambdapSI}
\end{eqnarray}
where the first sum in the right hand side in Eq. (\ref{eq:lambdapSI}) is taken over all $p$ spin sequences $\{k^{-}\}_{p-1}$ generated from the sequence $\{k\}_{p}$ removing  one spin (flipping it backwards) while the second summation is taken  over all $N-p$ spin sequences $\{k^{+}\}_{p+1}$ generated from the sequence $\{k\}_{p}$ flipping one additional spin. The forward approximation corresponds to the assumptions 
 \begin{eqnarray}
 {\rm Re} ~ \lambda_{\{k\}_{p}} \approx \frac{\Gamma}{(E-\Phi_{\{k\}_{p}})}\sum  {\rm Re} ~\lambda_{\{k^{-}\}_{p-1}},
 \nonumber\\
{\rm Im} ~ \lambda_{\{k\}_{p}} \approx \frac{\Gamma}{(E-\Phi_{\{k\}_{p}})} \sum {\rm Im} ~ \lambda_{\{k^{+}\}_{p+1}}. 
\label{eq:lambdaForwpSI}
\end{eqnarray}
Using Eq. (\ref{eq:GFEq9SI}) one can evaluate ${\rm Re} ~ \lambda_{\{k\}_{p}}$ as 
 \begin{eqnarray}
 {\rm Re} ~ \lambda_{\{k\}_{p}} \approx \sum_{\{t\}}\frac{\Gamma^{p-1}}{(E-\Phi_{t_{1}})(E-\Phi_{t_{1}t_{2}})...(E-\Phi_{t_{1}t_{2}...t_{p}})}
\label{eq:lambdaForpASI}
\end{eqnarray}
where the sum is taken over all $p!$ permutations $\{t\}$ of the given sequence $\{k\}_{p}$. 

Equations for the functions $\lambda$ for $n$ spin sequences $\{k\}_{n}$ can be formally  solved using the diagonal Green function $G_{\{k\}_{n}}$ for the states $a_{\{k\}n}$ similarly to Eq. (\ref{eq:GFEq7}) as 
 \begin{eqnarray}
 \lambda_{\{k\}_{n}}= G_{\{k\}_{n}}\sum \lambda_{\{k^{-}\}_{n-1}}. 
\label{eq:GFnForwFinSI}
\end{eqnarray}
Assume that  ${\rm Im}~ \lambda_{\{k\}_{n}} \approx {\rm Im}~ G_{\{k\}_{n}}\sum {\rm Re}~\lambda_{\{k^{-}\}_{n-1}}$ similar to Eq. (\ref{eq:lambdaForwpSI}). Using Eq. (\ref{eq:lambdaForwpSI}) to calculate ${\rm Im}~ \lambda_{k}$ for a single flipped spin and Eq. (\ref{eq:lambdaForpASI}) for the real part of functions $\lambda$   one can obtain the self-consistent equation in the $n^{th}$  order forward approximation generalizing the second order approximation, Eq. (\ref{eq:AbouChacraMod1SI}), as ($R_{a}=E-\Phi_{a}$) 
  \begin{eqnarray}
{\rm Im} ~\Sigma_{a} = \Gamma^{2}\sum_{\{k\}}\left[ D^{(n)}(\{\epsilon\}_{k}, \{U\}_{k}, R_{a})\right]^2
\frac{{\rm Im}~ \Sigma_{\{k\}}}{(E-\Phi_{a_{\{k\}}})^2}, 
\nonumber\\
 D^{(n)}(\{\epsilon\}, \{U\}, R) 
\nonumber\\
=\sum_{\{t\}}
\left[\prod_{i=1}^{n-1}\frac{\Gamma}{R-\sum_{j=1}^{i}\epsilon_{t_{j}}+\sum_{j' < j \leq i}U_{t_{j} t_{j'}}}\right],
\label{eq:AbouChacraModnSI}
\end{eqnarray}
where the sum in the first equation is taken over all ${N}\choose{n}$ sequences $\{k\}_{n}$ of $n$ spins, while the second equation introduces the kernel function $D$ as the sum taken over all $n!$ permutations $\{t\}$ of the given sequence $\{k\}$. The sequences $\{\epsilon\}_{k}$, $\{U\}_{k}$ represent transition energies and binary interactions for spins belonging to the sequence $k$. 

The structure of the $n^{th}$ order term is illustrated in Fig. \ref{fig:nord}. The structure of the kernel function is identical to the $n^{th}$ order forward approximation \cite{Laumann14,Laumann16,Antonello16}.

Eq. (\ref{eq:AbouChacraModnSI}) can be analyzed using the same approach $<e^{-s{\rm Im}\Sigma}>=1-s^{\beta}A$ as used in Ref. \cite{AbouChacra73}.  Then one can rewrite the equation determining the localization threshold in the form 
\begin{eqnarray}
A(R) 
={N \choose n} \Gamma^{2\beta}\prod_{i=1}^{n}\int_{l} p(\epsilon_{i})d\epsilon_{i} \prod_{j<i}\int_{-\infty}^{\infty}g(U_{ij})dU_{ij}  
\nonumber\\
\times \left| D^{(n)}(\{\epsilon\}, \{U\}, R)\right|^{2\beta}
\frac{A(R-\sum_{i=1}^{n}\epsilon_{i}+\sum_{i<j}U_{ij})}{|R-\sum_{i=1}^{n}\epsilon_{i}+\sum_{i<j}U_{ij})|^{2\beta}},
\label{eq:AbouChacraModn}
\end{eqnarray}
where integration domains $l$ are chosen to cutoff singularities in all denominators  similarly to Eq. (\ref{eq:AbouChacra2}), and the function $g(x)$ is a Gaussian distribution  with a zero average and root mean squared $x$ equal to $4J/\sqrt{N}$. 

The case $\beta=1/2$ is examined below. This is consistent with the standard forward approximation and can be justified in the domain of the forward approximation given by Eq. (\ref{eq:ForwAppl2}) using the methods of Ref. \cite{AbouChacra73}. 

Each of $n!$ terms contributing to the kernel function $D$ in Eq. (\ref{eq:AbouChacraModn})  corresponds  to a permutation $\{t\}$ of the sequence $1, .. n$. It  contains an $n^{th}$ order singularity in the denominator, realized at spin flip energies given by 
\begin{eqnarray}
\epsilon_{t_{1}}=0, ~ \epsilon_{t_{2}}=U_{t_{1}t_{2}}, ~ \epsilon_{t_{3}}=U_{t_{1}t_{3}}+U_{t_{2}t_{3}},... 
\nonumber\\
\epsilon_{t_{n}}=U_{t_{1}t_{n}}+...+U_{t_{n-1}t_{n}}.
\label{eq:Singularn}
\end{eqnarray}
The sum of $n$ terms $U_{ij}$ is of order $U_{n} \sim J\sqrt{n/N}$ and therefore  all spin flip energies $\epsilon_{i}$ near singular points belong to the domain $-U_{n}\leq \epsilon_{i} \leq U_{n}$. Outside that domain the integrals over energies rapidly converge provided that $\beta \approx 1/2$. This can be easily seen for $n=2$ where the function $D$ can be expressed as $D=\Gamma U_{ij}/(\epsilon_{i}(\epsilon_{i}+U_{ij}))$ and the integral over $\epsilon_{i}$ converges at $\epsilon_{j} \sim U_{ij}$. The more general statement for $n>2$  can be proved using the mathematical induction method  so one can restrict the integration domains for integrals over energies $\epsilon_{i}$ in Eq. (\ref{eq:AbouChacraModn}) to $-U_{n}\leq \epsilon_{i} \leq U_{n}$ within the logarithmic accuracy. 

These integrals diverge logarithmically and the restrictions to the integration domains should be applied following Refs. \cite{Anderson58,AbouChacra73}. Since the characteristic maximum energy is given by the inter-spin interaction $U_{n}$ the integration domain for each factor in the denominators of the kernel function $D$ in Eq. (\ref{eq:AbouChacraModn}) representing the multiple spin flip energies  is chosen with the logarithmic accuracy as $(-U_{n}, -\Gamma^2/U_{n}) \cup (\Gamma^2/U_{n}, U_{n})$. In other words these energies should not approach exact resonances (Eq. (\ref{eq:Singularn})) closer than the constraint energy $\Gamma^2/U_{n}$ (cf.  Refs. \cite{Anderson58,AbouChacra73}). The prefactors in the definitions of integration domains and cutoffs are not very significant since all dependencies on them are expressed in logarithmic form.

Further simplification of Eq. (\ref{eq:AbouChacraModn}) can be attained by setting $R=0$ and making the resonant approximation  as in Eq. (\ref{eq:AbouChacra3}) so the factors $A(0)$ can be omitted in both sides. This leads to the equation  determining  the localization threshold in the form  
\begin{eqnarray}
1=X_{n}
=4{N \choose n} \ln\left(\frac{U_{n}}{\Gamma}\right)\Gamma\prod_{i=1}^{n}\int_{l} p(\epsilon_{i})d\epsilon_{i}
\nonumber\\
\times \left[\prod_{j<i}\int_{-\infty}^{\infty}g(U_{ij})dU_{ij}  \right]
\nonumber\\
\times\left| D^{(n)}(\{\epsilon\}, \{U\}, R)\right|
\delta\left(\sum_{i=1}^{n}\epsilon_{i}-\sum_{j=1}^{n}\sum_{i=1}^{j-1}U_{ij}\right). 
\label{eq:AbouChacraModn1}
\end{eqnarray}

\subsubsection{Applicability of a forward approximation}
\label{subsec:ForwSI}

Using typical forward approximation parameters one can examine the relevance of the forward approximation at large $n \gg 1$. For instance consider the applicability of Eq. (\ref{eq:lambdaForpASI}) which essentially determines the kernel function in Eq. (\ref{eq:AbouChacraModnSI}). This equation has been obtained skipping contribution of sequences with extra spin in the first line of Eq. (\ref{eq:lambdaForwpSI}) 
 \begin{eqnarray}
 \frac{\Gamma}{(E-\Phi_{\{k\}_{p}})}\sum  {\rm Re} ~\lambda_{\{k^{+}\}_{p+1}}.
\label{eq:lambdaForwpCorrSI}
\end{eqnarray}
This contribution can be estimated using  Eq. (\ref{eq:lambdaForpASI}) for longer sequences. According to the above analysis only extra spins with flip energies $\epsilon \leq U_{n}$ should be included and there are $N_{n}=NU_{n}/W_{1}$ of such spins.  The sum in the correction, Eq. (\ref{eq:lambdaForwpCorrSI}), contains $(n+1)!N_{n}$ terms compared to the main contribution in Eq. (\ref{eq:lambdaForpASI}) leading to the forward approximation but each term contains an extra factor $\Gamma^2/(U_{n}^2)$. Thus the correction term differs from the leading approximation by the factor $(N\Gamma/W_{1})\cdot (n\Gamma/U_{n})$. Since in the transition point one has $\Gamma \approx W_{1}/N$ the correction can be estimated as $(n\Gamma/U_{n})$ and the forward approximation is applicable under the condition 
\begin{eqnarray}
n <  \left(\frac{U}{\Gamma}\right)^2. 
\label{eq:ForwAppl2}
\end{eqnarray}
A similar constraint can be derived considering equations for imaginary parts (second line in Eq. (\ref{eq:lambdaForpASI})). 

Eq. (\ref{eq:ForwAppl2}) can be interpreted qualitatively as comparing the pure forward processes (spin sequences) involving only one flip of each spin  with the processes involving multiple flips of each spin as suggested in the main text.

\subsubsection{Estimate of a localization threshold}

Since the terms in the sum determining the kernel function $D$ possess Levy statistics in the absence of the constraint for minimum denominators, one can evaluate the integrals there assuming that this is the case and then verify the validity of the assumption about the statistics. In the case of Levy statistics the sum is determined by the maximum term and one can replace the absolute value of the sum  with the sum of absolute values. Then the integrals in Eq. (\ref{eq:AbouChacraModn1}) can  be evaluated  within the logarithmic accuracy as (integrals over interaction constants can be evaluated replacing these constants in the argument of logarithm by their typical value $U$) 
\begin{eqnarray}
X_{n}=\frac{N!}{(N-n)!}\Gamma^{n}p(0)^{n}2^{2n} \ln\left(\frac{U\sqrt{n}}{\Gamma}\right)^{n}.  
\label{eq:AbouChacraModn2}
\end{eqnarray} 
The localization-delocalization transition should take place at $X_{n}=1$. 

The validity of this result can be tested examining the validity of the Levy statistics for the sum of $n!$ singular terms in the kernel function $D$. The absolute value of the denominator in each term is distributed nearly uniformly within the domain $(0, U_{n})$. Then the minimum value of denominator $d_{n}$ can be estimated considering the probability that each out of $n$ factor there exceeds $d_{n}$. This probability is given by $(1-d_{n}/U_{n})^{n}$ so one gets $d_{n} \sim U_{n}/n$. If $d_{n} < \Gamma^2/U_{n}$ the lower constraint for the denominators becomes significant  and the assumption about Levy statistics fails. This takes place simultaneously with the failure of the forward approximation, Eq. (\ref{eq:ForwAppl2}), so Eq. (\ref{eq:AbouChacraModn2}) is applicable until the forward approximation is valid. This result is used in the main body of the manuscript.

\subsection{Localization threshold in the case of a weak interaction} 
\label{sec:SIWeak}

To estimate the localization threshold in the case of a weak interaction one can use an approximate matching Bethe lattice problem for the lattice containing $N$ spins in a locally diagonalized representation. Each product state couples to $N^2/2$ states different from the given state by two spin flips. The ``diagonal"  energy difference $E_{ij}$ of initial and final states different by flips of spins $i$ and $j$ is determined by the spin energy difference $E_{i}\pm E_{j} = \sqrt{\epsilon_{i}^2+\Gamma^2} \pm  \sqrt{\epsilon_{j}^2+\Gamma^2}$. Only one half of spin transitions corresponding to the negative sign can be resonant, corresponding to flip-flop transitions. The coupling matrix element responsible for the flip-flop  transition is given by $4J_{ij}\Gamma^2/(E_{i}E_{j})$. 

One can apply  the self-consistent theory of localization to this problem similarly to the previous considerations considering the contribution of resonant transitions with arbitrary energies $E_{i} \approx E_{j}= E  > \Gamma$. The integration domain in the ``logarithmic" approximation is given by $U\Gamma^2/E^2 <|E_{i}-E_{j}| < U$ leading to the logarithmic factor $\ln(E/\Gamma)$ which can be large for large energies $E > \Gamma$. The problem in this form matches the Bethe lattice problem with varying coupling strengths (due to different energies $E$ involved) and a localization criterion can be obtained integrating these contributions together in the form
\begin{eqnarray}
1=2N^2 p(0)^2\Gamma_{c}^2\left<\left|J_{ij}\right|\right>\int_{-\infty}^{\infty}d\epsilon \frac{1}{\epsilon^2+\Gamma^2} \ln\left(\frac{\epsilon^2+\Gamma^2}{\Gamma^2}\right) 
\nonumber\\
\approx 7 N^2 p(0)^2 J\Gamma_{c}/\sqrt{N}. 
\label{eq:WekIntSI}
\end{eqnarray}
The integral over energies is determined by $\epsilon \sim 4.5 \Gamma$ (the domain $(-4.5\Gamma, 4.5\Gamma)$ gives around half of the total integral) which gives some qualitative justification for logarithmic accuracy requiring the argument of the logarithm to be much greater than $1$.

Evaluating integrals in Eq. (\ref{eq:WekIntAnsSI}) one can estimate the localization threshold as 
\begin{eqnarray}
\Gamma_{c}=\frac{\eta_{1}}{7 N^{\frac{3}{2}} p(0)^2 J}, 
\label{eq:WekIntAnsSI}
\end{eqnarray}
where the factor of $\eta_{1} \sim 1$ accounts for the possible inaccuracy of the logarithmic approximation in Eq. (\ref{eq:WekIntSI}). This is a conservative estimate in the sense that in addition to the logarithmic contribution there is some constant contribution from the domain $|E_{i}-E_{j}| > U$, which is ignored in Eq. (\ref{eq:WekIntSI}). Consequently  the localization threshold can be overestimated. The localization threshold estimate, Eq. (\ref{eq:WekIntAnsSI}), is used in the main body of the manuscript.

\end{document}